\documentclass{PoS}
\usepackage{graphicx}

\newcommand\POWHEG{{\tt POWHEG}}
\newcommand\BOX{{\tt POWHEG-BOX}}
\newcommand\HERWIG{{\tt HERWIG}}
\newcommand\HWpp{{\tt HERWIG++}}
\newcommand\PYTHIA{{\tt PYTHIA}}
\newcommand\MCatNLO{{\tt MC@NLO}}

\def\({\left(} 
\def\){\right)}

\title{Single-top production with the POWHEG method} 

\ShortTitle{Single-top production with the POWHEG method}

\author{\speaker{Emanuele Re}
  \thanks{Preprint numbers: IPPP/10/46, DCPT/10/92.}\\
  IPPP, Durham University\\
  E-mail: \email{emanuele.re@durham.ac.uk}}

\abstract{We describe briefly the \POWHEG{} method and present results
  for single-top s- and t-channel production at hadron colliders.}

\FullConference{XVIII International Workshop on Deep-Inelastic Scattering and Related Subjects\\
  April 19 -23, 2010\\
  Convitto della Calza, Firenze, Italy}

\begin{document}

\section{The POWHEG method}
Monte Carlo event generators and NLO QCD calculations are important to
give reliable predictions for signals and backgrounds relevant for
collider phenomenology.  To exploit the advantages of both the
approaches it is clear that a method to include NLO QCD corrections to
event generators is desirable, especially in view of the amount of
data that LHC will collect.
The \POWHEG{} method is a prescription to interface NLO calculations
with parton shower generators avoiding double counting of emissions,
achieving therefore the aforementioned task.

The method was first suggested in ref.~\cite{Nason:2004rx}, and was
described in great detail in ref.~\cite{Frixione:2007vw}. It has been
applied to several processes at
lepton~\cite{LatundeDada:2006gx,LatundeDada:2008bv} and
hadron~\cite{Nason:2006hfa,Frixione:2007nw,Alioli:2008gx,Hamilton:2008pd,Papaefstathiou:2009sr,Alioli:2008tz,Hamilton:2009za,Alioli:2009je,Nason:2009ai}
colliders.  Unlike \MCatNLO{}~\cite{Frixione:2002ik}, \POWHEG{}
produces events with positive weight, and, furthermore, does not
depend on the subsequent shower Monte Carlo program: it has been
interfaced to \HERWIG{}~\cite{Corcella:2000bw},
\PYTHIA{}~\cite{Sjostrand:2006za} and \HWpp{}~\cite{Bahr:2008pv}.  Up
to now all the implementations have been tested against the ones
available in \MCatNLO{}: reasonable agreement has been found as well
as the reason for the (few) differences encountered (see for
example~\cite{Nason:2010ap} for a recent discussion).

In the \POWHEG{} formalism, the generation of the hardest emission is
performed first, according to the distribution given by
\begin{equation}
\label{eq:master}
d\sigma=\bar{B}\left(\Phi_{B}\right)\,d\Phi_{B}\,\left[\Delta_{R}\left(p_{T}^{\min}\right)+
\frac{R\left(\Phi_{R}\right)}{B\left(\Phi_{B}\right)}\,\Delta_{R}\left(k_{T}\left(\Phi_{R}\right)\right)\,d\Phi_{\mathrm{rad}}\right]\,,
\end{equation}
where $B\left(\Phi_{B}\right)$ is the leading order contribution,
\begin{equation}
\label{eq:bbar}
\bar{B}\left(\Phi_{B}\right)=B\left(\Phi_{B}\right)+
\left[V\left(\Phi_{B}\right)+\int d\Phi_{\mathrm{rad}}\, R\left(\Phi_{R}\right)\right]
\end{equation}
is the NLO differential cross section integrated on the radiation
variables while keeping the Born kinematics fixed
($V\left(\Phi_{B}\right)$ and $R\left(\Phi_{R}\right)$ stand
respectively for the virtual and the real corrections to the Born
process), and
\begin{equation}
\Delta_{R}\left(p_{T}\right)=\exp\left[-\int d\Phi_{\mathrm{rad}}\,\frac{R\left(\Phi_{R}\right)}{B\left(\Phi_{B}\right)}\,\theta\left(k_{T}\left(\Phi_{R}\right)-p_{T}\right)\right]\,
\end{equation}
is the \POWHEG{} Sudakov. With $k_T\left(\Phi_{R}\right)$ we denote
the transverse momentum of the emitted particle. The cancellation of
soft and collinear singularities is understood in the expression
within the square bracket in eq.~(\ref{eq:bbar}).
Partonic events with hardest emission generated according to
eq.~(\ref{eq:master}) are then showered with a $k_T$-veto on following
emissions.\footnote{We refer to~\cite{Nason:2004rx,Frixione:2007vw}
for all the technicalities that we are neglecting here.}

Eq.~(\ref{eq:master}) is useful to understand easily the main
properties of the method. Nevertheless, it is not a trivial task to
build a code that takes into account all the technicalities (and the
numerical subtleties) hided in eq.~(\ref{eq:master}).
Although it has been possible to build standalone codes for simple
processes,
an automated tool (named \BOX{}~\cite{Alioli:2010xd}) was build in
order to deal with more complex cases. This program has been already
used to implement Higgs-boson production via vector-boson fusion and
V + $1$jet~\cite{POWHEG_Zjet} with \POWHEG{}.  Recently, some of the
previous implementations have been also included in this
package.
\section{Single-top with POWHEG}
Single-top processes are the ones where only one top quark is produced
in the final state. In literature it is customary to categorize the
production modes as $s$-, $t$- and $Wt$-channel, in accordance with
the virtuality of the $W$ boson involved in the scattering.  Cross
sections are smaller than the $t\bar{t}$ pair one, due to their weak
nature. This fact, together with the presence of large W + jet and
$t\bar{t}$ backgrounds, makes the single-top observation very
challenging, so that this signal has been observed only recently by
the CDF and D0 collaborations~\cite{Abazov:2009pa,Aaltonen:2009jj}.

Despite of its relative small cross section, single-top production is
an important signal for several reasons (for a recent review see for
example ref.~\cite{Bernreuther:2008ju}) and it will be studied
thoroughly at the LHC.  It is therefore natural to provide the
experimental community with a tool that performs the merging between
NLO and parton showers for this process.  This has been accomplished
some time ago within the \MCatNLO{}
framework~\cite{Frixione:2005vw,Frixione:2008yi}, whereas a \POWHEG{}
implementation for $s$- and $t$-channel has been completed more
recently in a standalone code, and then also included in the \BOX{}
package.\footnote{Results presented in the following have been
  obtained with the \BOX{} package.}  An implementation for
$Wt$-channel is also under way.

\begin{figure}[!htb]
\begin{center}
\includegraphics[width=0.8\textwidth]{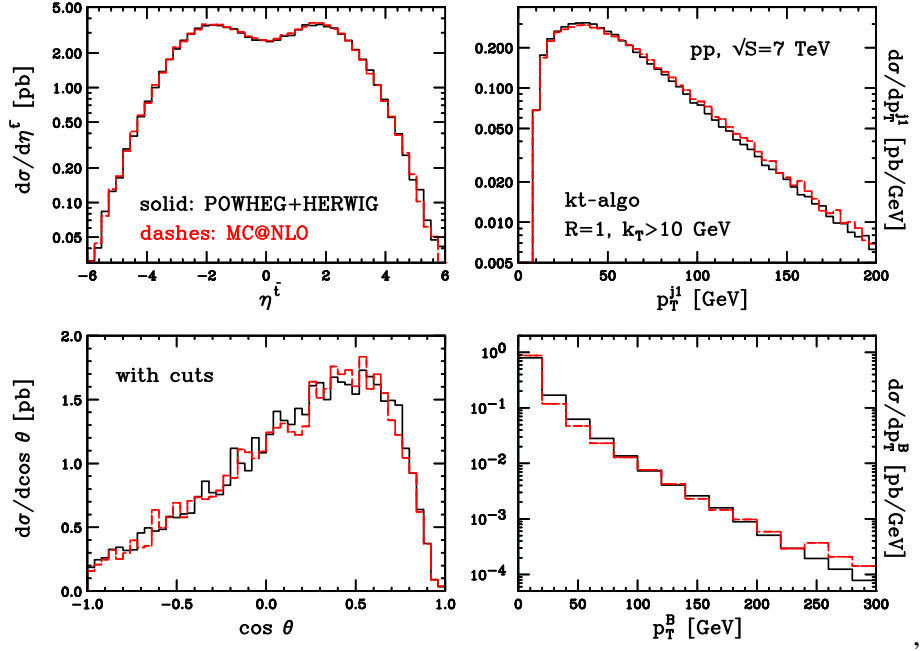},
\end{center}
\caption{Comparisons between \POWHEG{} and \MCatNLO{} for $t$-channel
  single-top $\bar{t}$ production at the LHC. Cuts used for angular
  correlation are: $|\eta_b|<2.5$, $p_T^b>50$ GeV, $2.5<|\eta_{sj}|<5.0$,
  $p_T^{sj}>30$ GeV, $|\eta_{\ell}|<2.5$, $p_T^{\ell}>30$ GeV,
  $E_T^{miss}>20$ GeV. With $sj$ we denote the spectator jet.}
\label{fig:tbar_POW_MCNLO}
\end{figure}

\begin{figure}[!htb]
\begin{center}
\includegraphics[width=0.8\textwidth]{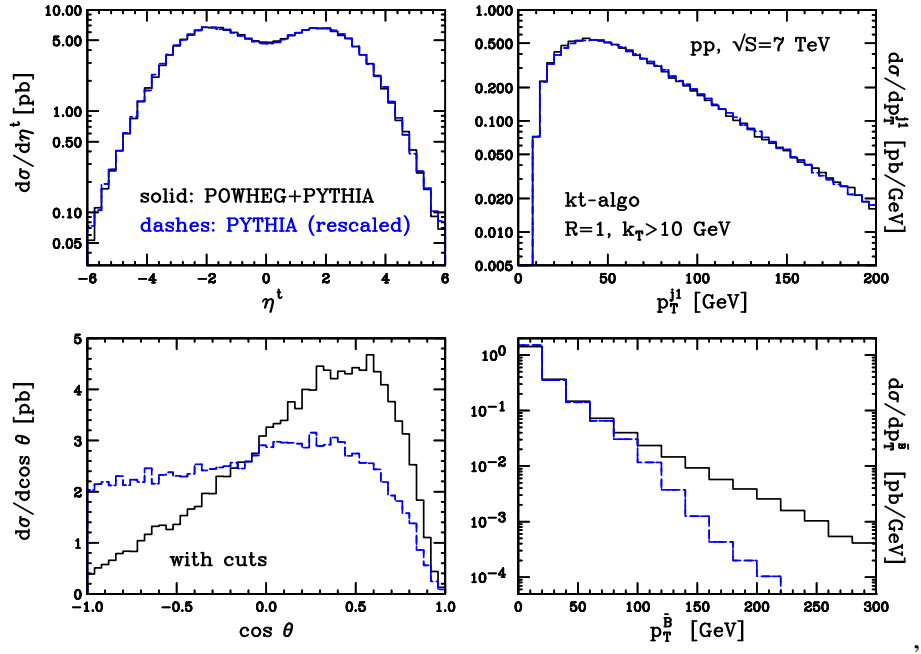},
\end{center}
\caption{Comparisons between \POWHEG{} and \PYTHIA{} for $t$-channel
  single-top $t$ production at the LHC. For angular correlation the
  same cuts as above have been used.}
\label{fig:t_POW_PYT}
\end{figure}

In figures~\ref{fig:tbar_POW_MCNLO} and \ref{fig:t_POW_PYT} we present
some typical distributions for single-top $t$-channel $\bar{t}$ and
$t$ production at the LHC respectively. Results have been obtained
choosing $\sqrt{S}=7$ TeV, $m_t=173.1$ GeV and $m_W=80.4$ GeV. As
PDF's we used the CTEQ6M set. Plots have been obtained including the
top-quark semileptonic decay, but removing the branching ratio.

The very good agreement with the \MCatNLO{} results shows that the
method works properly.  In the following we discuss in more detail
the differences with respect to \PYTHIA{} predictions.

We start by recalling that \PYTHIA{} results have been rescaled to the
\POWHEG{} (and \MCatNLO{}) ones, since their normalization is only LO
accurate.


Concerning the $p_T$ spectrum of the hardest $\bar{b}$-flavoured
hadron ($b$-flavoured for anti-top production), the difference is due
to the absence of matrix-element corrections in \PYTHIA{}: the $g\to
b\bar{b}$ splitting (typical of $t$-channel single-top) is accurate
only in the collinear limit (being it performed by the shower) and
therefore the high-$p_T$ tail of \PYTHIA{} result shows a lack of
events.

In the plot showing the angular correlation between the direction of
the charged lepton coming from the top decay and the spectator jet
(i.e.~the hardest non $b$-flavoured jet),\footnote{Here we use the
  \emph{spectator basis}~\cite{Mahlon:1999gz} and we plot the angle as
  seen in the top-quark rest frame.} \POWHEG{} and \MCatNLO{} are in
good agreement but differ remarkably with \PYTHIA{}.  This
distribution measures spin-correlation effects between the production
and the decay process~\cite{Mahlon:1999gz}. To take into account these
effects, in \POWHEG{} the top decay is generated with a procedure
similar to the one used in \MCatNLO{}~\cite{Frixione:2007zp}.  Instead
in \PYTHIA{} top decay-products are generated uniformly and the
corresponding angular distribution results flat.\footnote{We notice
  that the damping at high values of $\cos\theta$ is due to the cuts
  applied.}

\section*{Acknowledgements}
Results presented in this proceeding have been obtained in
collaboration with S. Alioli, P. Nason and C. Oleari.  The author
wants also to acknowledge G. Corcella for the invitation to the
conference and the opportunity to give the talk remotely.

\end{document}